\documentstyle [twoside,11pt,draft]{article}
\begin{document}
\input epsf
\setcounter{page}{1}
%

\Large
\begin{center}
{\Large \bf AMS, a particle spectrometer in space
\footnote{Talk given at the XXIV Symposium on Nuclear Physics, January 3-6, 2001, 
Taxco, Mexico. }}
\vspace{0.3cm}
\par
{\large M. Bu\'enerd} \\
\par
\vspace{0.2cm}
{\small\it Institut des Sciences Nucl\'{e}aires, IN2P3/CNRS, 
53 av. des Martyrs, 38026 Grenoble cedex, France} \\
\par
{\small for the AMS collaboration} 
\end{center}
\vspace{0.5cm}
\parbox{6.3in}{\small\underline{Abstract}: The results of the precursor flight of the AMS 
experiment are reviewed, the interpretation of the measured proton flux reported, and
the prospects for the forthcoming main phase on the International Space Station outlined.
}
\normalsize
\section {Introduction}
Accurate measurements of particle fluxes close to earth have been 
performed recently by the AMS experiment, bringing a body of excellent new data on the 
particle populations in the low altitude terrestrial environment. These results should 
rejuvenate the long standing interest of a broad community of scientists for the interactions 
between the cosmic ray (CR) flux and the atmosphere and for the dynamics of particles in the 
earth neighborhood. They certainly open new prospects for accurate studies of these 
phenomena to investigate the interaction mechanisms generating the observed populations.
\par
The AMS experiment took its first data during a precursor flight on june 2-12, 1998, on the
Space Shuttle DISCOVERY. The flight was primitively intended as a qualification test for 
the spectrometer instrumentation. The orbit altitude was close to 370 km. During 100 Hours 
of counting, about 10$^8$ events were recorded providing new results of high quality on 
the particle distributions at the altitude of the detector. Some of these results were rather 
unexpected. They illustrate the discovery potential of the experiment in its future steps.
\par
This contribution is devoted to a general presentation of the project, of the results 
obtained during this first experimental test and of their interpretation, and of the goals 
and plans of the forthcoming phase II of the experimental program. 
The first part will deal with a description of the measurements performed and questions 
raised on the dynamics of the detected particles in the earth environment, by the results 
obtained. The second part will describe a phenomenological approach based on a simulation 
to account for the observed distributions. The third and last part will consist of a 
description of the phase II AMS spectrometer which will begin on the International Space 
Station as of next october 2003, which will be very different from the version flown on the 
shuttle, and of its physics program. 
\par
%
\section {The AMS01 precursor flight}
%
The spectrometer operation during the flight has been very successful with only a few 
instrumental defects, having no significant consequence on the quality of the measurements 
achieved. 
\subsection {The spectrometer}
%
\begin{figure}[htb]
\begin{center}
\hspace{1cm}
\epsfysize=8cm
\epsfbox{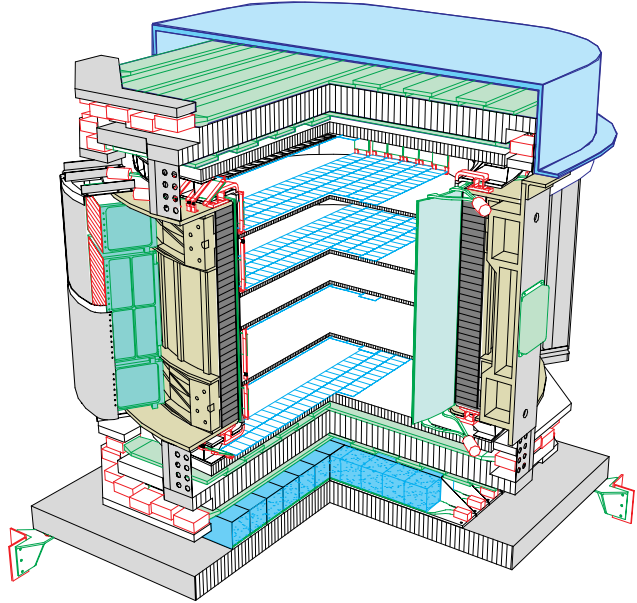}
\end{center}
\vspace{-0.5cm}
\caption{\small\it Open view of the AMS phase 1 spectrometer. 
}
\label{AMS01}
\end{figure}
Figure~\ref{AMS01} shows a cut view in perspective of the spectrometer which was flown on 
the shuttle. The apparatus included a cylindrical permanent magnet generating a 0.15~Tesla 
dipole field perpendicular to the axis of the cylinder inside its volume \cite{AIMANT}. 
The inner volume was mapped with a tracker consisting of 6 planes of silicon microstrips 
partially equipped at this stage, allowing the reconstruction of particle trajectories
\cite{TRACK} . The tracker planes also provided dE/dX measurements of the particles. 
Above and below the magnet, two double planes of scintillator hodoscopes with perpendicular 
orientations of their paddles, provided both a measurement of the particle time of flight 
(TOF) and of their specific energy loss (dE/dX). The paddle location and the position
sensitivity inside the paddles also provided a complementary determination of the particle 
hit coordinates, useful for background rejection. A skirt of scintillators around the 
tracker was used to veto on particles outside the fiducial angular acceptance of the 
counter. At the bottom of the device a threshold Cherenkov counter equipped with n=1.035 
aerogel material allowed $p/e^+$ and $\bar{p}$/e$^-$ discrimination below the $p(\bar{p})$ 
threshold around 4~GeV/c particles \cite{ATC}. 
\par
\subsection{Results}
Some of the results have already been published \cite{HEBAR,PROT1,PROT2,LEPT,HE}. 
The measured data are still under analysis however, and the physics issues addressed by the 
experiment are being actively investigated. Some of the latters are discussed on in the 
following. 
\subsubsection{Search for antimatter}
The first claimed objective of the experiment is the search for primordial antimatter in 
space. It was then very important to investigate the capability of the spectrometer to 
identify antiparticles with Z$\geq$2, and to identify and reject background events. \\
\par
$\bullet$ {\bf Antihelium {\normalsize\cite{HEBAR}} - } 
%
\begin{figure}[htb]
\begin{center}
\hspace{0.5cm}
\epsfysize=8cm
\epsfbox{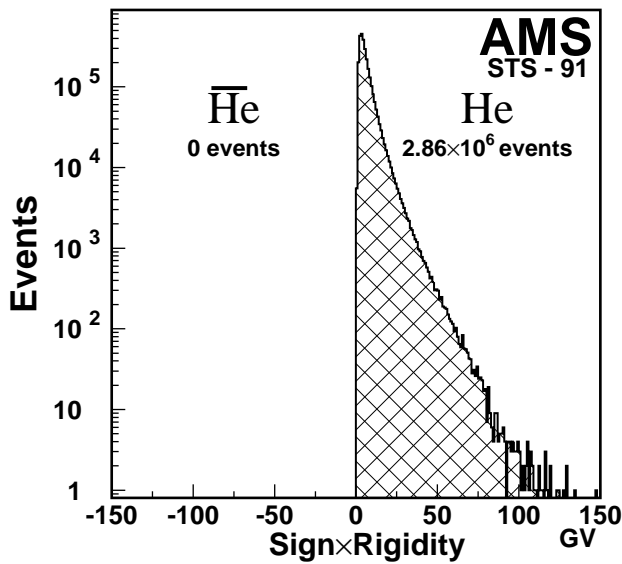} 
\end{center}
\vspace{-0.5cm}
\caption{\small\it Distribution of rigidity times sign of charge for |Z=2| particles. 
Positive(resp negative) values correspond to $He$(resp $\overline{He}$) particles.
}
\label{AHE}
\end{figure}
Figure~\ref{AHE} shows the spectral distribution of Z=2 particles as a function of their 
rigidity, i.e., momentum/charge, the sign of the charge being measured by the sign of the 
trajectory curvature in the tracker. Positive rigidities correspond to He particles, whereas 
antiHeliums are expected on the negative side. A few fake antiheliums due to soft 
interactions in the detector, were rejected by means of appropriate cuts on the energy 
deposit in the tracker planes. 
\par
Finally the experiment has allowed to set a new lower limit on the $\overline{He}$/$He$ 
fraction in cosmic rays, of 1.1 10$^{-6}$. See \cite{BESS} for recent results from the BESS 
experiment. \\
\par
$\bullet$ {\bf Antimatter nuclei Z$>$2 - } 
The particle identification capabilities of the spectrometer have been used to search also 
for antimatter nuclei with Z$>$2. This search has been negative so far. The limit obtained 
will be reported in a future publication.
\subsubsection{Protons {\normalsize\cite{PROT1,PROT2}}}
\par
\begin{figure}[htb]
\begin{center}
\vspace{2cm}
\epsfxsize=15cm
\epsfbox{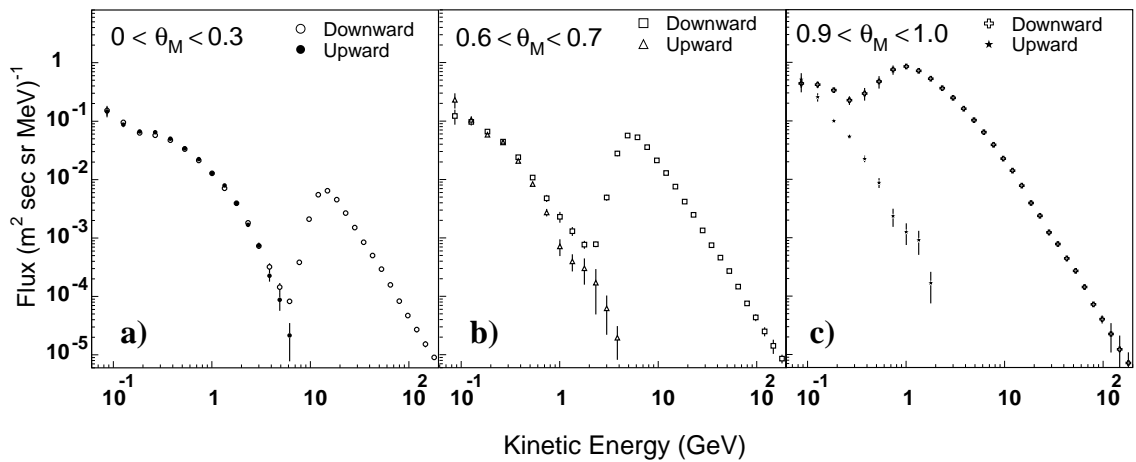}
\end{center}
\vspace{-0.5cm}
\caption{\small\it Downgoing and upgoing proton distributions measured by AMS at variou for 
different bins of latitudes.
}
\label{PROTONS}
\end{figure}
The CR proton distribution was already very well known from previous experiments before the 
AMS flight. The measurements were intended to be used for checking and calibrating the 
experiments, no new result being expected. 
Figure~\ref{PROTONS} show the kinetic energy distributions of incoming particles (towards 
earth) measured by AMS in bins of latitude. The spectra show some expected features like 
the power law decrease with energy. The geomagnetic cutoff (GC) due to the sweeping away of 
particles by the earth magnetic field below a critical momentum, is clearly observed in the
spectra, decreasing from about 15~GeV around the equator down to zero in the polar region.
The spectrum at high latitudes is in good agreement with previous measurements. Although no 
significant flux was expected below GC, a strong rise of the spectra at low energy is 
observed at all low latitudes with a strong enhancement in the equatorial region.  
The Albedo (outgoing particles) spectra at the same latitude do not show as expected the 
high energy features due to the incoming CR flux, but they display one single component 
peaked at low energy and overlapping almost perfectly (to within 1\%) with the low E 
component of the incoming flux. These features indicate that we are dealing with a 
population of trapped particles circling around earth magnetic field lines, exactly 
as in the Van Allen belts but at much higher energy and much closer to earth. 
This will be confirmed by the analysis reported below. 
%
\subsubsection{Leptons {\normalsize \cite{LEPT}}}
%
\begin{figure}[htb]
\begin{center}
\hspace{-1.5cm}
\epsfysize=8cm
\epsfbox{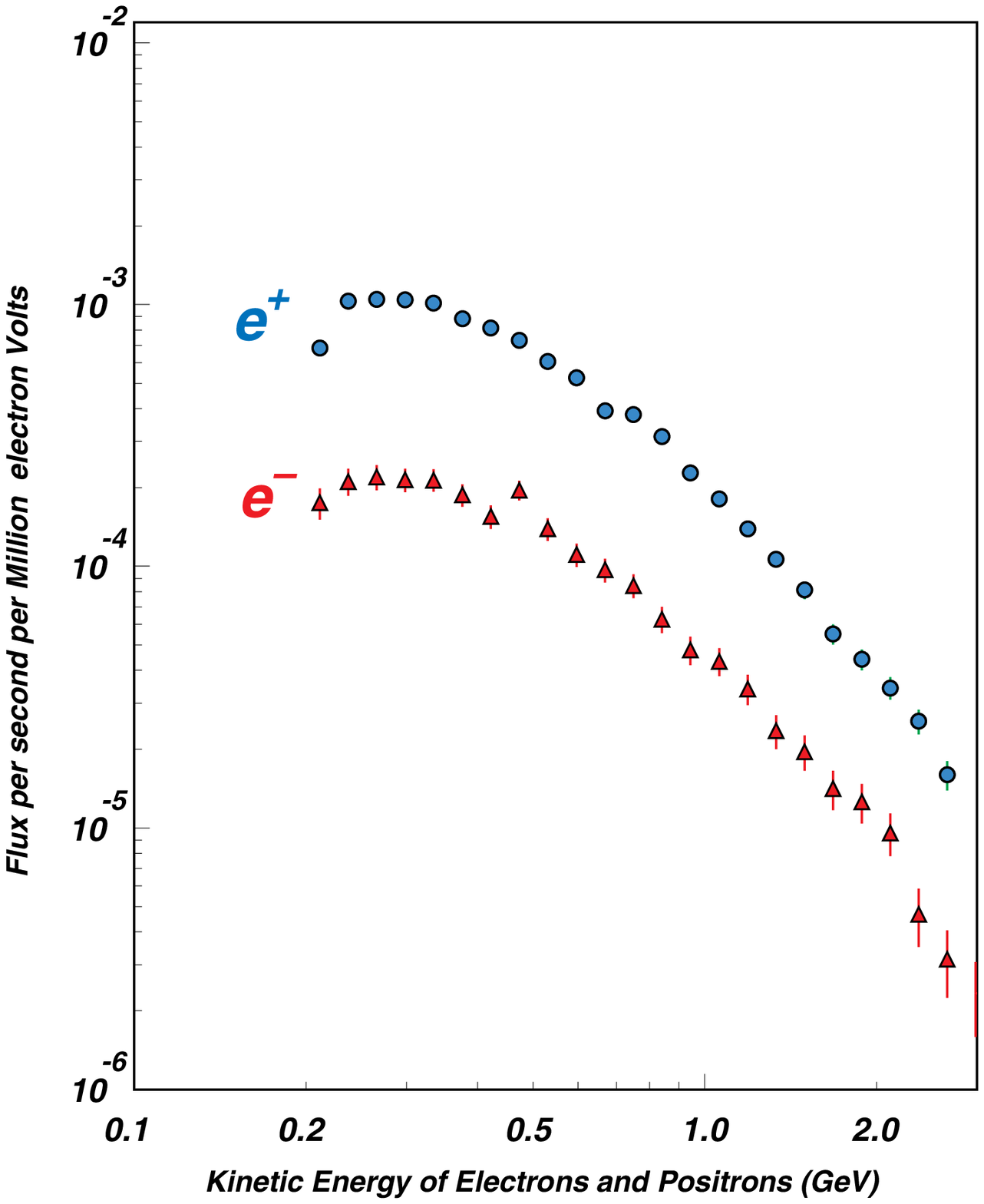} 
\epsfysize=8cm
\epsfbox{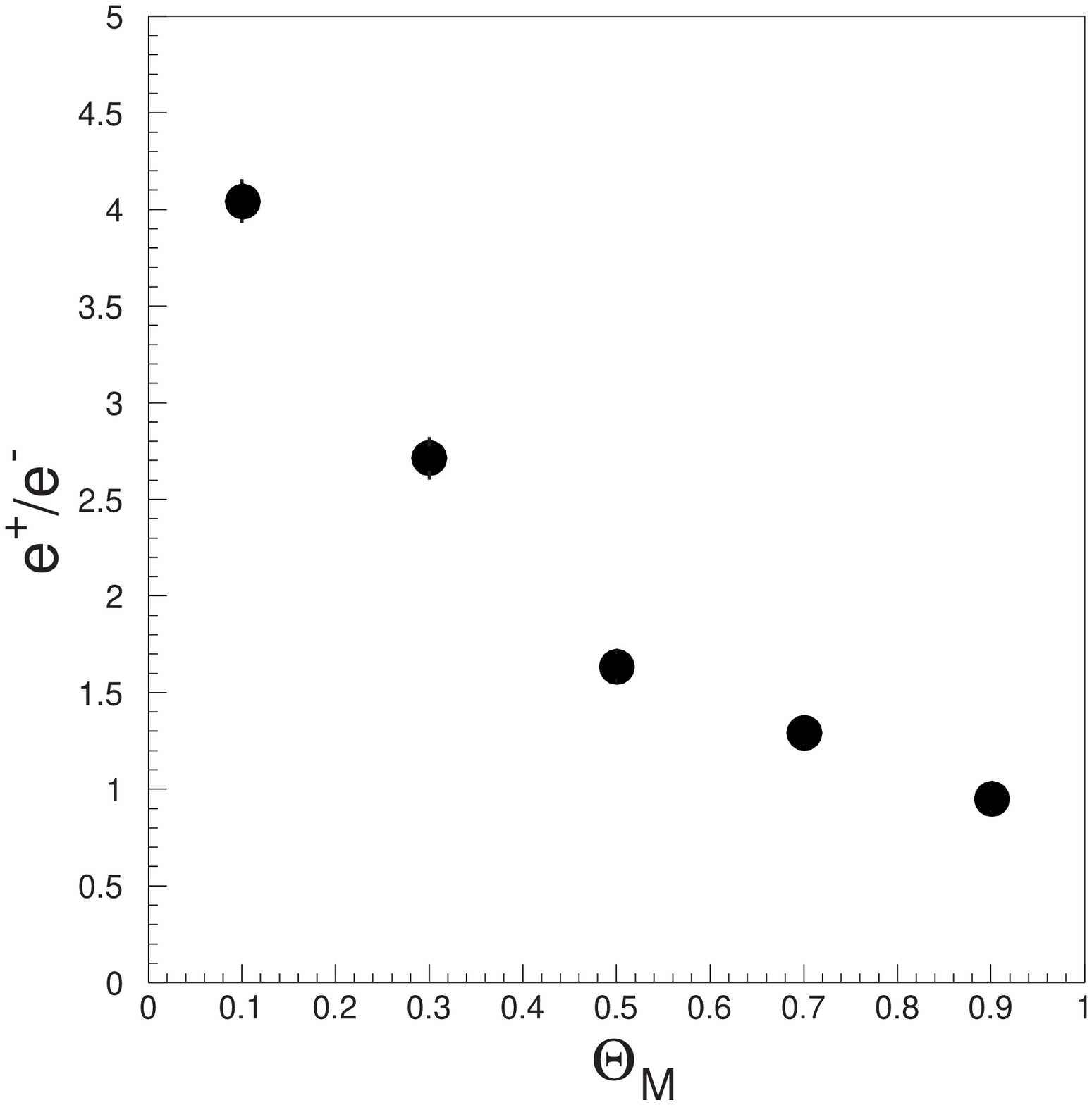}  
\end{center}
\vspace{-0.5cm}
\caption{\small\it  Left: Electron and positron spectra measured by AMS in the equatorial 
region. Right: Distribution of the experimental positron over electron ratio versus 
latitude of the measurement.
}
\label{LEPTONS}
\end{figure}
The flux of leptons has been measured up to about 100~GeV for electrons. It was limited to 
about 3 GeV for positrons by the $p/e^+$ discrimination range set by the Cherenkov counter 
threshold for protons.
\par
$\bullet$ {\bf Electrons}
The electron spectra show quite similar features as the proton spectra, with the low energy 
component of the downgoing flux and the upgoing flux almost perfectly overlapping in the 
equatorial region. In addition these components of the lepton flux have exactly the same 
shape to within statistical errors, as for protons, indicating that the particles are 
likely involved in the same dynamical process.  
\par   
$\bullet$ {\bf Positrons}
The positron spectra are similar to the electron's over the range investigated. The 
surprising feature is that the positron to electron flux ratio is about 4 in the equator 
region, while in the cosmic flux it is about 0.1, and about one in the atmosphere. The
origin of this feature is an open question which is being addressed by the groups of the 
collaboration.  
\par
Figure~\ref{LEPTONS} shows the distributions of electrons and positrons over the positron 
ID range in the equatorial region (left) and the distribution of the e$^+$/e$^-$ ratio in
latitude. 
\subsubsection{Ions}
\hspace{0.4cm} $\bullet$ {\bf Deuterium {\normalsize\cite{DEUT}} - }
The flux of deuterium has been measured and some preliminary results are available. 
\par
$\bullet$ {\bf Helium {\normalsize\cite{HE}} - } 
%
\begin{figure}[htb]
\begin{center}
\hspace{1.5cm}
\epsfysize=8cm
\epsfbox{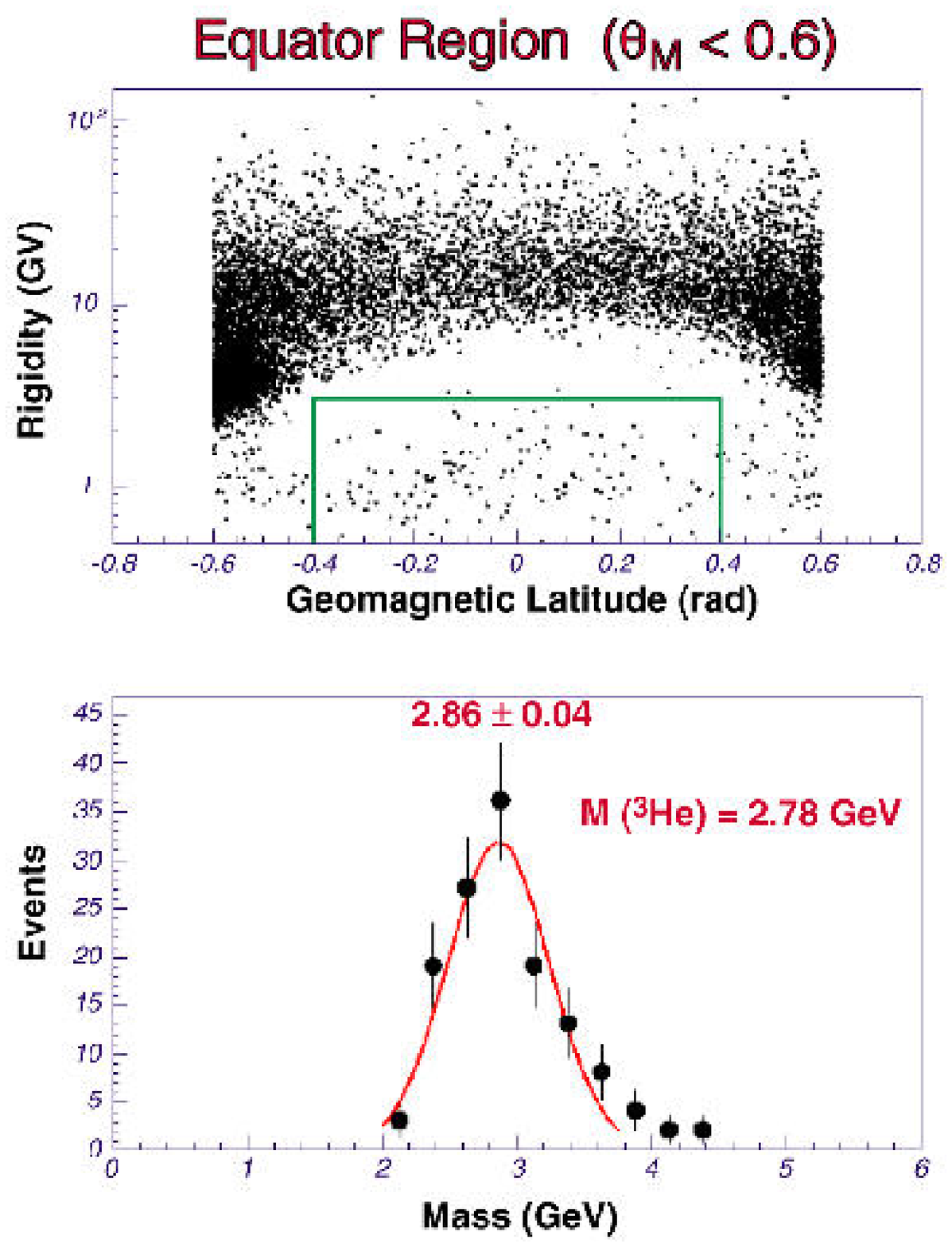} 
\end{center}
\vspace{-0.5cm}
\caption{\small\it Top: Scatter plot of the rigidity versus latitude for Z=2 particles, 
showing the low rigidity limit set by the geomagnetic cutoff. Note the small population 
observed below the cutoff. Bottom: Reconstructed mass of the events inside the rectangle
above. The mean value is found to be close to the mass of the $^3$He isotope. 
}
\label{HELIUMS}
\end{figure}
The measured flux of Helium is in agreement with previous measurements and doesn't show 
a strong rise of flux below GC as the proton flux does. However a small flux of $^3$He is 
found below GC, which originates probably at least partly from the fragmentation of cosmic 
$^4$He (figure~\ref{HELIUMS}. A consistent picture of these population based on known 
nuclear reaction mechanisms is being invetsigated to account for these populations of light 
nuclei \cite{DERHE}.
\par
$\bullet$ {\bf Z$>$2 Nuclei - }
Some significant samples of light ions with 2$<$Z$\leq\approx$10 have been measured during 
this run. They are still being analyzed. 
\section {Origin of the measured proton flux {\normalsize\cite{DER00}}} 
%
\subsection {Simulation program}
The inclusive spectrum of protons at the altitude of AMS (390-400km) has been calculated 
by means of a computer simulation program built to this purpose. 
CR particles are generated with their natural abundance and momentum distributions. They 
are propagated inside the earth magnetic field. Particles are allowed to 
interact with atmospheric nuclei and produce secondary protons with cross sections and 
multiplicities as discussed below. Each secondary proton is then propagated and allowed to 
collide as in the previous step. 
A reaction cascade can thus develop through the atmosphere. The reaction products are 
counted when they cross the virtual sphere at the altitude of the AMS spectrometer, upward 
and downward. Particles undergo energy loss by ionisation before and after the interaction. 
Multiple scattering effects have not been included at this stage. Each event is 
propagated until the particle disappears by either colliding with a nucleus, or being 
stopped in the atmosphere, or escaping to outer space beyond twice the production altitude. 
Note that particles are counted each time they cross the sphere of detection altitude.  
The contributions of trapped particles are thus weighted statistically with their numbers 
of crossings, which increases their contribution to the final spectrum. 
\par
The secondary nucleon spectrum generated has to cover two orders of magnitudes in kinetic 
energy, between about 100~MeV and 10~GeV. The main component of proton production cross 
section was obtained by means of analytical relations fitted to 14.6~GeV $p+Be$ data. 
The scaling properties of the cross section have been checked with the FRITIOF/PYTHIA 
(L\"und) event generator. Since this generator is not expected to account for the very low 
energy and backward proton emission (target-like to negative rapidities), this latter 
component was incorporated using a parametrization. The respective 
contributions to the total multiplicity-weighted proton production cross-section, were 
352~mb for the QE component and 88~mb for the DI components. Cross sections on atmospheric 
nuclei were renormalized from the original data or parametrizations obtained on different 
nuclei, using ratios of geometrical cross sections. 
\subsection {Results}
\begin{figure}[hbt]
\begin{center}
\hspace{0.5cm}
\epsfysize=8cm
\epsfbox{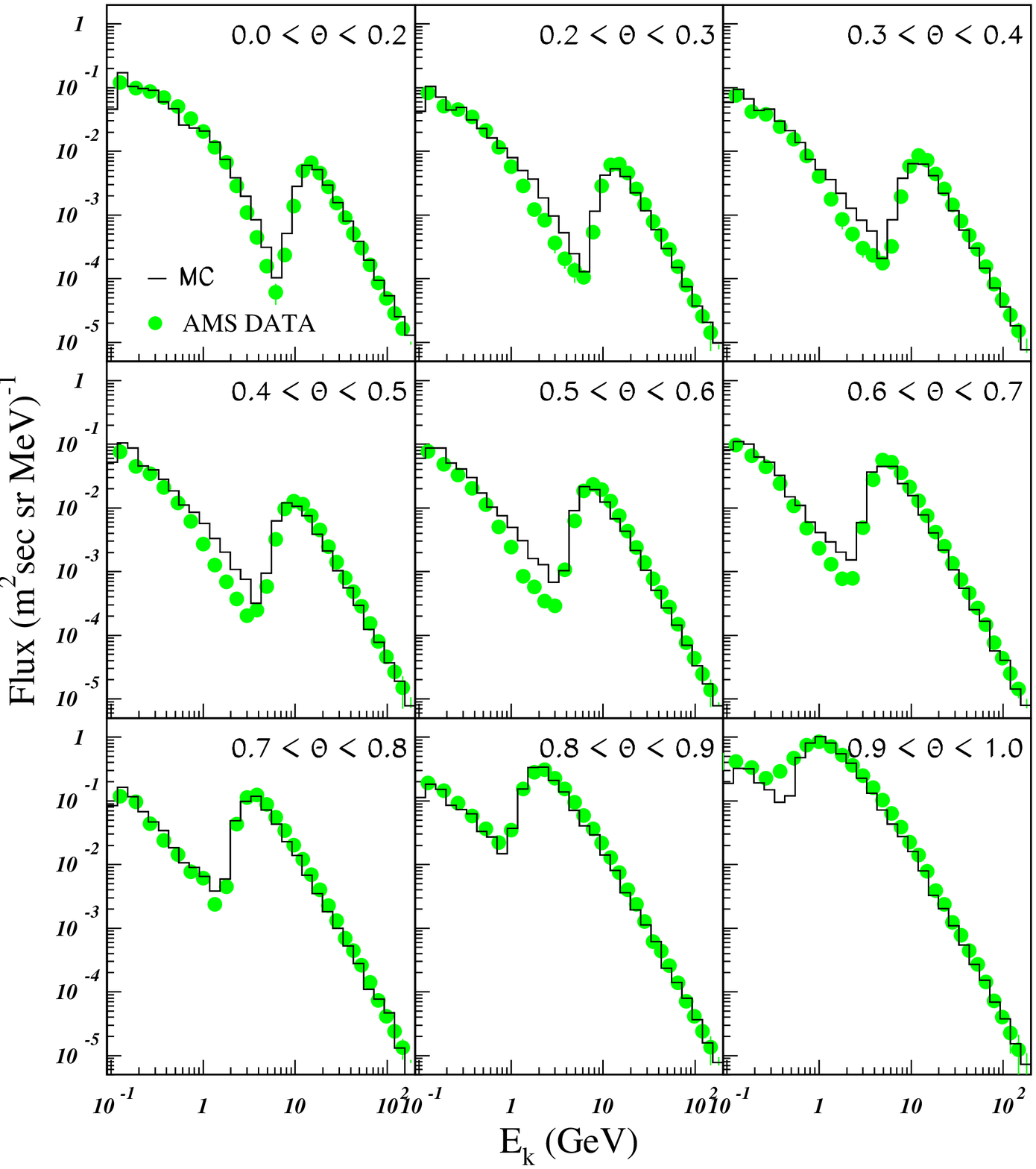} 
\hspace{0.2cm}
\epsfysize=8cm
\epsfbox{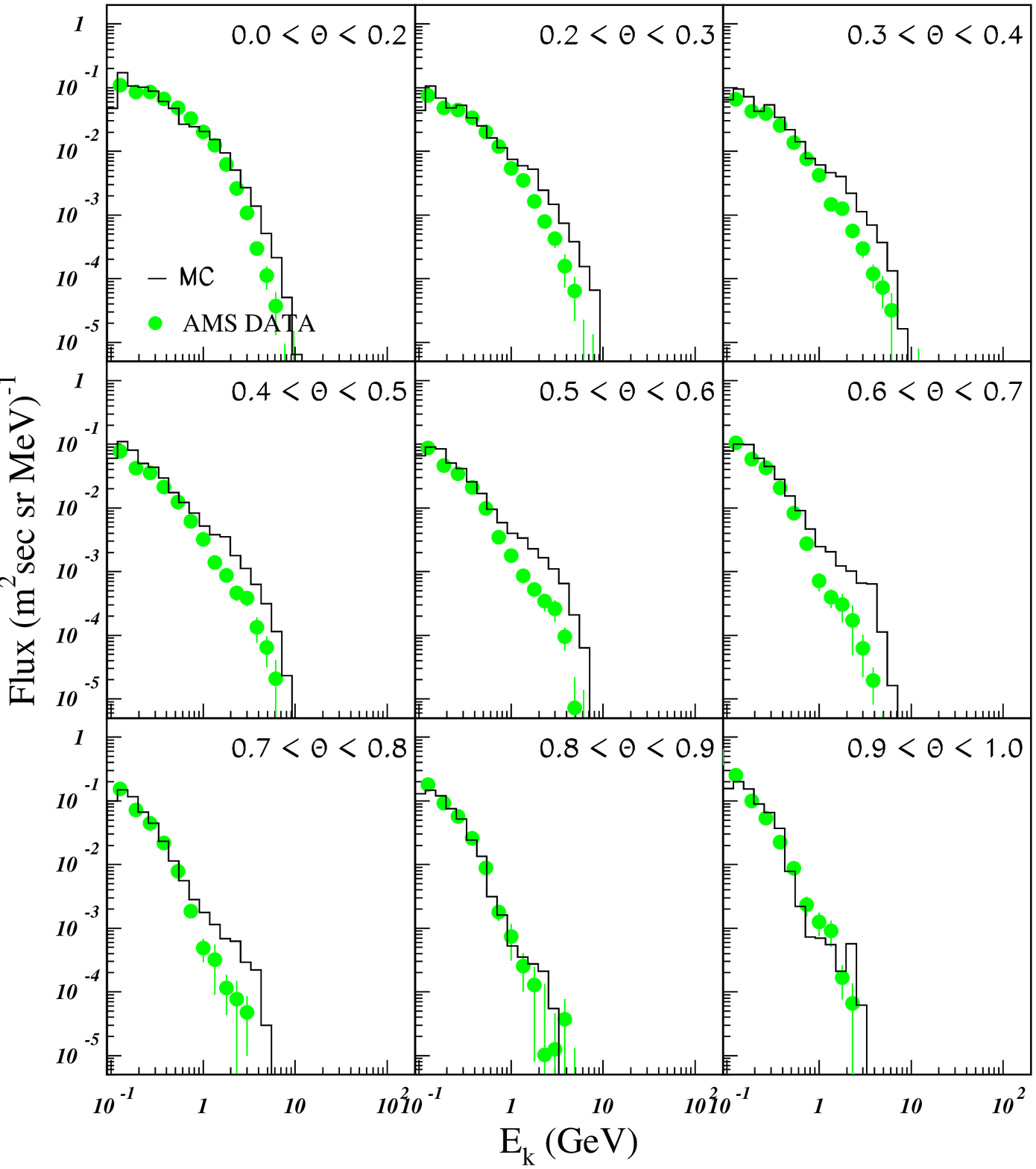} 
\end{center}
\caption{\small\it Experimental kinetic energy distributions \cite{PROT1} for 
bins in latitudes (full circles), compared to the results of the simulation (full line)
described in the text, for downward (left) and upward (right) protons.}
\label{DISTRIB}
\end{figure}

Many features of the dynamics of particles in the earth magnetic field appear in the 
simulation results. Some of them are discussed in the published paper. Others will be
reported later.
Figure~\ref{DISTRIB} shows the experimental kinetic energy distributions of downward 
(left) and upward (right) protons measured for some bins of latitude, compared to the 
results of the simulation. No free parameter was used for normalization to the data:  
The calculated results are entirely determined by the physics input to the calculation. 
It is seen that the agreement between the data and the simulation result is remarkably
good, at all latitudes and for both the inward and outward flux. In particular, the cutoff 
region is particularly well reproduced, which indicates that the processing of the particle 
dynamics and kinematics is good. The shaded histograms in the figure correspond to secondary 
particles in the simulation. The fraction of events originating from the DI component of 
the proton production cross section described previously, vary from about 10\% in the 
equatorial region up to 25\% in the polar region, with a momentum distribution peaking 
at low kinetic energy and distributed below 500~MeV. 
\par
It can be concluded from this result that the proton flux measured by AMS can be accounted 
for to a good accuracy by the single interaction of the incoming CR flux with the
atmosphere. 
%
%
\subsection {Future prospects}
This successful result opens a world of new prospects on the phenomenology of particles in
the earth environment. Beside the ongoing investigations of the other AMS results described 
above, some other issues of general or particular interest are being addressed or will be 
addressed soon, like the study of the atmospheric neutrino flux and the secondary antiproton 
populations close to earth. The same type of approach can be used also for particle 
propagation in the galactic interstellar medium and the study of the various astrophysical 
issues associated to this propagation.  
\par
\vspace{1cm}
\section {AMS02, a particle observatory in space}
%
\begin{figure}[htb]
\begin{center}
\hspace{1cm}
\epsfysize=14cm
\epsfbox{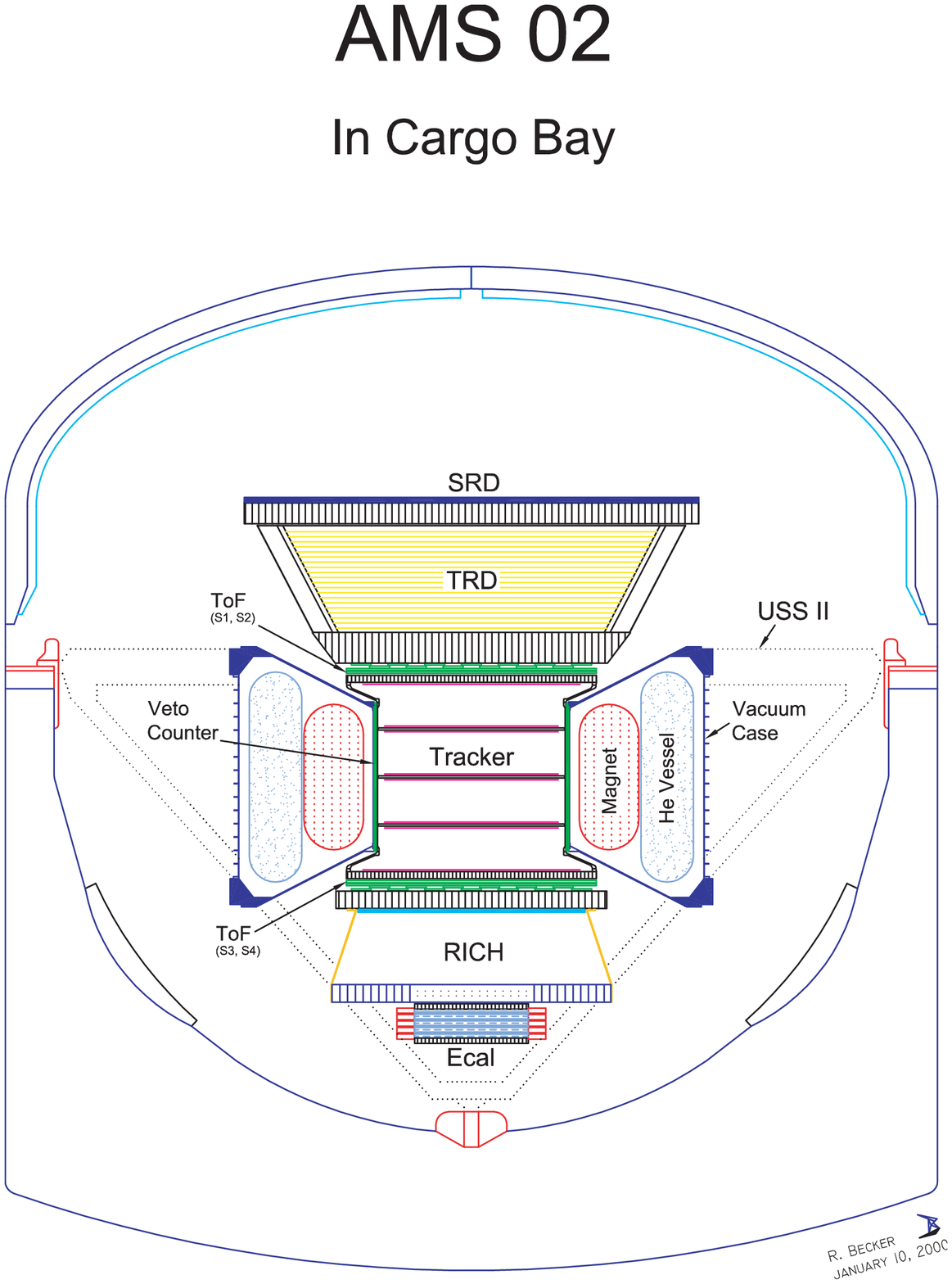} 
\end{center}
\vspace{-0.5cm} 
\caption{\small\it  Schematic view of the AMS02 spectrometer architecture on its support 
structure (USS) in the Space Shuttle bay.
}
\label{AMS2}
\end{figure}
The second phase of the AMS experiment will begin on october 2003 with the launch of the 
spectrometer and its installation on the International Space station (ISS) for a 3 to 5 years
campaign of measurements. 
\subsection {Spectrometer structure}
The new spectrometer shown on figure~\ref{AMS2}, will improve on the AMS01 
(figure~\ref{AMS01}) version by many respects. Its main change consists of the permanent 
magnet (B$_{max}$=0.15~T) of AMS01 being replaced by a superconducting magnet in AMS02 
(B$_{max}$=1~T) which will result in a 6 times better resolution and 6 times larger momentum 
range, because of this larger magnetic field. In addition, several new detectors 
will be implemented in AMS02: A transition radiation detector (TRD) will provide lepton 
identification up to above 300 GeV and an improved tracking accuracy. A Cherenkov imager 
(RICH) will allow nuclear isotope identification up to about 13~GeV/c for mass around 
carbon \cite{RICH}. An electromagnetic calorimeter (ECAL) will provide the energy 
measurement of electromagnetic particles $\gamma$, leptons, and their discrimination from 
hadrons up to the TeV range. The synchrotron radiation detector (SRD) would provide
e$^+$/e$^-$ identification/discrimination at very high energies \cite{SRD}.
\subsection {Physics program}
The physics program to be covered with the new instrument is wide, with a high discovery 
potential and a significant probability of unexpected results and new findings. Basically
the spectrometer will be able to accumulate statistics larger by 3 to 4 orders of magnitudes
than those measured so far by other embarked experiments, for all the species studied. 
The range in rigidity will extend from around 300~MV up to 3~TV, depending on the particle 
species, with a good identification capability for leptons, hadrons, and ions, provided by 
the spectrometer instrumentation. The TRD counter will provide lepton-hadron (mainly 
positron-proton) discrimination up to the TRD proton threshold around 300~GeV. The RICH 
will allow both charge and mass measurements: The charge can be obtained from the photon
yield, proportionnal to Z$^2$, for momenta going from threshold up to the upper limit of the 
counter momentum resolution. The mass measurement is obtained from the ring image processing, 
over a range limited by the space resolution of the counter as shown in the table (see 
\cite{RICH} for details). The final performances and range will depend on the choice of 
Cherenkov radiators. The electromagnetic calorimeter will allow the energy measurement 
of electromagnetic particles, leptons and photons, up to the TeV range, and lepton-hadron
discrimination from the measured P/E ratio over the same range.
\par  
%
\begin{figure}[htbp]
\begin{center}
\vspace{-1cm}
\hspace{1cm}
\epsfysize=9cm
\epsfbox{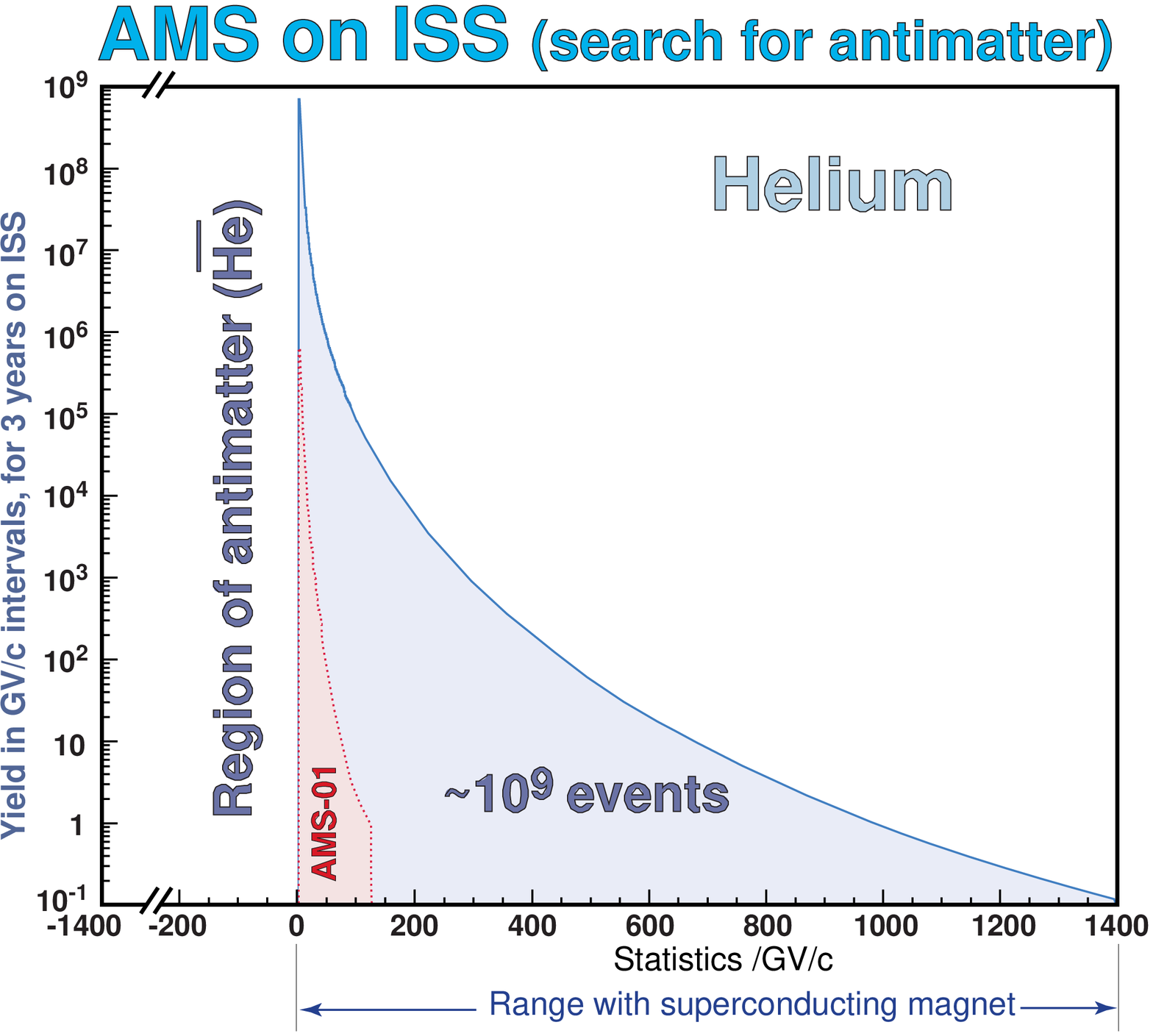} 
\end{center}
\vspace{-0.5cm}
\caption{\small\it Comparison between the $He$ statistics obtained in AMS01
and expected in AMS02, relevant to the search for $\overline{He}$ in the cosmic flux.
}
\label{AHE2}
\begin{center}
\vspace{2cm}
\hspace{1cm}
\epsfysize=9cm
\epsfbox{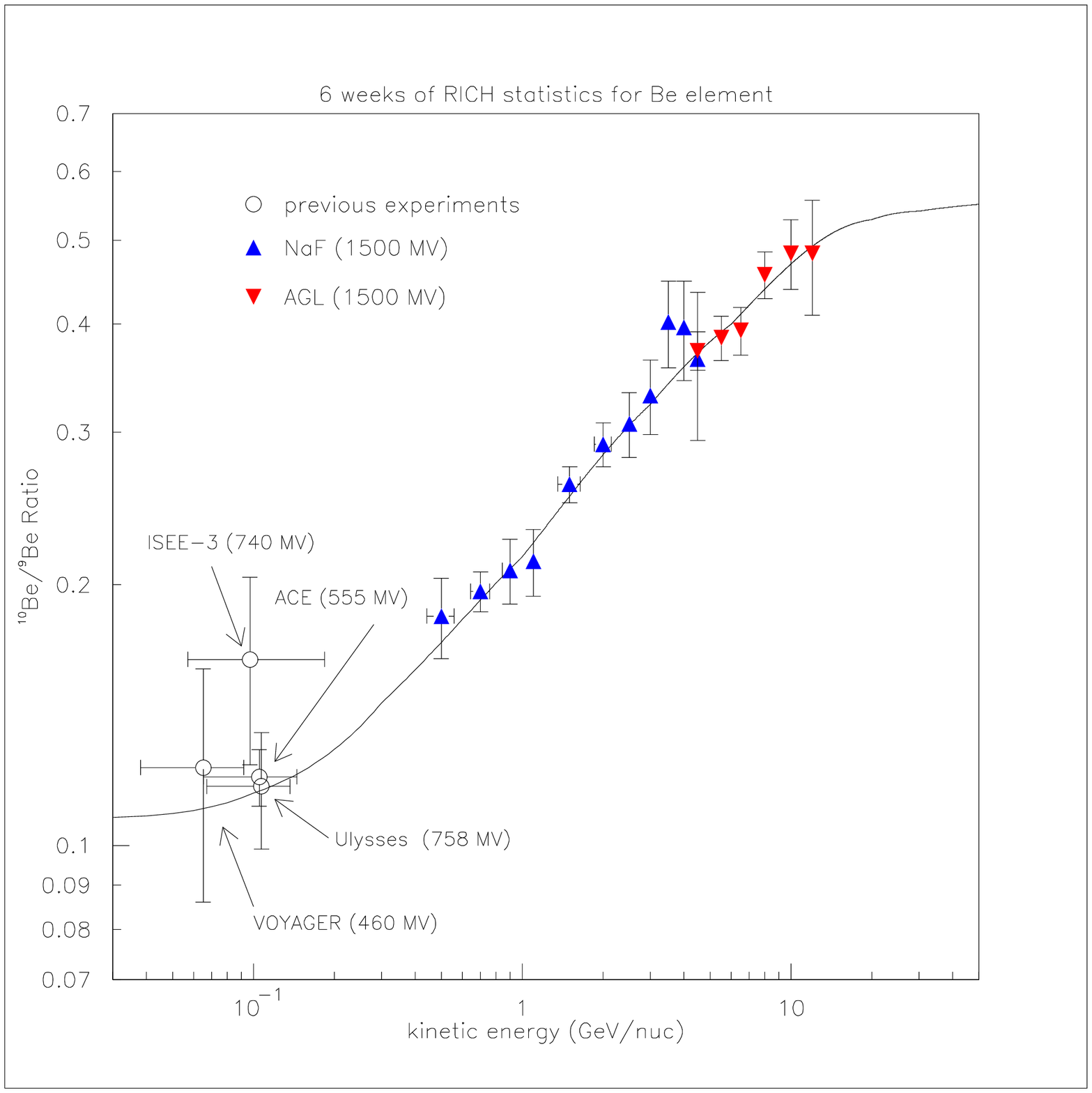} 
\end{center}
\vspace{-0.5cm}
\caption{\small\it Expected statistics for the $^{10}$Be isotope for 6 weeks of counting 
from a simulation using the RICH of AMS.
}
\label{BE10}
\end{figure}
\par
These features will allow a high statistics study of many cosmic ray species including $e^+$,
 $e^-$, $p$, $\bar{p}$, and the lightest ions $d,t,^{3,4}$He. Heavier light ions will also 
be studied with mass identification up to A$\approx 20$, and elements up to Z$\approx 20$ 
depending on the final performances of the instrumentation of the spectrometer (RICH in 
particular).
Unstable ions like $^{10}$Be, and $^{26}$Al are of particular interest since they provide
a measurement of the time of confinement of charged particles in the galaxy (galactic 
chronometers) \cite{BO00}. 
The corresponding antimatter nuclei will be searched with equivalent instrumental 
performances in identification and kinematic range. 
The metrological perspectives are summarized in table~\ref{PROSP}.
\begin{table}
\label{CHARAC}
\begin{center}
\caption{\it Summary of the particle detection and identification range of the AMS02 
 spectrometer. The upper instrumental limits are set either by the momentum measurement 
 accuracy (at the highest momenta) or by the range of identification of the particle. The 
 lower values are set by the low momentum cutoff of the magnetic spectrometer or by the 
 range of particles in detectors, or by threshold (like Cherenkov) effects. Statistical 
 limits are ignored. The instrumental limits on the ranges for (nuclear) matter and 
 antimatter are the same. The given numbers should be considered as orders of magnitudes.
 The true limits will depend critically on the relative statistics of the particles to be 
 identified versus background particles, like e$^+$/$p$ or $\bar{p}$/$e^-$.  
 The momenta are given in GeV/c or GeV/c per nucleon when applicable.} 
\vspace{0.5cm}
\begin{tabular}{llll}
\hline\hline
 Particles        & P$_{min}$ & P$_{max}$ & Comments  \\
\hline 
e$^-$             & $\approx$0.3 & $\approx$3000 & Upper limit set by rigidity resolution \\
e$^+$             & $\approx$0.3 & $\approx$300  & Upper limit set by TRD \\
proton            & $\approx$0.3 & $\approx$3000 & Upper limit set by rigidity resolution \\
\hline
\multicolumn{4}{c}{\bf Charge identification of elements } \\
Ions Z<$\approx$20 & $\approx$0.3 & $\approx$1500 ? & Depending on RICH performances \\
\hline
\multicolumn{4}{c}{\bf Mass identification of isotopes} \\
Ions A<4             & 1 to 4  & $\approx$20 & Depending on RICH performances \\
Ions 4<A<$\approx$20 & 1 to 4  & $\approx$12 & \\
\hline
\multicolumn{4}{c}{\bf Antimatter } \\
$\bar{p}$         & $\approx$0.3 & $\approx$3000 & Depending on $\bar{p}/e^-$ discrimination\\
$\overline{ions}$ & $\approx$0.3 & $\approx$1500 & $\overline{He}$, $\overline{C}$ \\
\hline \hline 
\end{tabular}
\label{PROSP}
\end{center}

\end{table}
These capabilities will allow to address with an unmatched sensitivity the main scientific 
objectives of the program: \\
1) The search for antimatter in space \cite{MBAR}: The experimental signature for the 
detection of an antinucleus basically requires a determination of its charge in module and 
sign, the key point being the sign of the charge provided by the radius of curvature of the 
trajectory in the tracker, its accuracy and contamination by various backgrounds. The 
results discussed previously have shown that the background level is under control (see 
details in \cite{HEBAR}). It will be improved with the updated the tracker equipment in 
number of planes and readout electronics. \\
2) The search for dark matter in space through the signature of neutralino annihilations. 
The latter are expected to generate kinematic structures in the spectra of their annihilation
products \cite{DARKM}. Such structures will be searched for in $\bar{p}$ and e$^+$ spectra.  
3) In addition, the study of the various ions within the spectrometer (RICH) range of 
identification, will be achieved, in particular for the $^{10}$Be isotope. This is 
illustrated on figure~\ref{BE10} with the result of a simulation incorporating the 
ID resolution of the spectrometer provide by the RICH and a theoretical $^{10}$Be 
distribution in momentum. It is seen that the measured sample for 6 weeks of counting time
on the ISS would provide highly accurate data over a range totally unexplored so far 
by previous experiments. The study of this sample will provide an estimate of the accuracy
on the propagation parameters to be expected \cite{BO00}.
See ref~\cite{GAMMAS} for a study of a possible high energy gamma ray astronomy program with
AMS.
%
\section {SUMMARY and CONCLUSION}
In summary, it has been shown that the first engineering flight of the AMS experiment has 
been very successful, both instrumentally and scientifically. This first step has provided a 
significant number of new unexpected physics results on the particle populations close to 
Earth. Although no new physics has emerged so far from these data, they are new and important 
and expected to provide some significant improvements in our understanding of the particle 
flux in the environment close to earth at the completion of the study. 
These result obtained in a region extensively explored by previous balloon or satellite 
experiments, illustrate the discovery potential of the spectrometer due to its instrumental
characteristics: large geometrical acceptance, large momentum range and dynamics and 
particle identification capability, and to the long duration of the measurement campaign 
scheduled for the main phase of the experiment on the ISS, where close to 3 orders 
magnitude more statistics than in the first flight will be collected in much better 
instrumental conditions.
%

%
\end{document}